\documentclass[aps,prl,twocolumn,superscriptaddress,reprint]{revtex4-2}

\usepackage{amsmath}
\usepackage{graphicx}
\usepackage{xcolor}
\usepackage[colorlinks,urlcolor=blue,citecolor=blue,linkcolor=blue]{hyperref}

\begin{document}

\title{Wavefront Mapping for Absolute Atom Interferometry}

\author{Joseph Junca}
\affiliation{Time and Frequency Division, National Institute of Standards and Technology, Boulder, Colorado 80305, USA}
\affiliation{Department of Physics, University of Colorado, Boulder, Colorado 80309, USA}
\author{John Kitching}
\author{William McGehee}
\email[]{william.mcgehee@nist.gov}
\affiliation{Time and Frequency Division, National Institute of Standards and Technology, Boulder, Colorado 80305, USA}

\date{\today}

	\begin{abstract}
		\noindent Wavefront distortions are a leading source of systematic uncertainty in light-pulse atom interferometry, limiting absolute measurements of gravitational acceleration at the 30 nm/s$^2$ level. Here, we demonstrate {\it in situ} spatially resolved measurement of the interferometer phase in a Mach-Zehnder atom interferometer as a tool to characterize and correct wavefront bias. By introducing controllable curvature of the Raman light using an adjustable collimation retro-reflector, we show that the bias due to parabolic wavefront curvature can be measured with 1 mrad uncertainty and that finite-size corrections impact the measured phase curvature. This measurement process could be adopted in optimized atom interferometer gravimeters to reduce wavefront bias uncertainty below the nm/s$^2$ level. 
	\end{abstract}

\maketitle

% \section{Introduction}
{\it Introduction}---Light-pulse atom interferometry (LPAI)~\cite{croninAtomInterferometers2009,abendAtomInterferometryIts2019} is used to make precise measurements of inertial effects~\cite{Peters1999,Zhang2023,Savoie2018}, to determine the values of fundamental constants~\cite{Rosi2014,Morel2020}, and to probe physics beyond the standard model~\cite{dimopoulosTestingGeneralRelativity2007,Sabulsky2019,Asenbaum2020}. Applications in inertial sensing have been pursued for decades, as atom interferometry offers a path to high sensitivity measurements with low absolute uncertainty. LPAI gravimeters exemplify this trend, reaching stability levels matching the best cryogenic relative gravimeters~\cite{Zhang2023} and approaching the absolute uncertainty of falling corner-cube gravimeters of $\approx$ 20 nm/s$^2$~\cite{niebauerNewGenerationAbsolute1995,francisEuropeanComparisonAbsolute2013}. Fieldable LPAI gravimeters and gravity gradiometers could impact a broad range of geo-science applications including ground water monitoring, resource extraction, and measurements of volcanic activity~\cite{farahUndergroundOperationBest2014,geigerHighaccuracyInertialMeasurements2020,antoni-micollierDetectingVolcanoRelatedUnderground2022,narducciAdvancesFieldableAtom2022a}. 
%Development of transportable LPAI gravimeters could enable long-term, continuous monitoring of these systems. 

The lowest reported absolute uncertainties in LPAI gravimeters are approximately 30 nm/s$^2$~\cite{Peters1999,gillotLNESYRTEColdAtom2016,Freier2016}, limited by several effects including Coriolis phases, magnetic shifts, light shifts, and wavefront aberrations. Many of these biases can be characterized with low uncertainty or rejected using differential measurements~\cite{Durfee2006,Gauguet2008}. Wavefront non-ideality of the interferometer light can directly bias the interferometric phase due to unequal sampling of the wavefront phase as the atoms expand or move during the measurement sequence~\cite{gauguetCharacterizationLimitsColdatom2009a,gillotLNESYRTEColdAtom2016,Freier2016}. Significant wavefront biases can be generated by wavefront curvature at the 10~km scale, strained vacuum windows, and Gouy phases---requiring {\it in situ} characterization to realize the lowest absolute wavefront uncertainty. 

% Wavefront curvature of the interferometer beams is typically needed at the  estimated in LPAI gravimeters using component-level testing and is often the leading source of uncertainty. The wavefront of the interferometer light is the ruler used to measure the desired inertial quantity, and deviations from an ideal flat phase front are sampled by the atoms as they expand during the interferometer sequence, generating bias on the integrated atomic phase. For LPAI driven by light pulses with a parabolic wavefront distortion, the acceleration bias depends only on the gas temperature and the radius of curvature of the optical wavefront. This bias is independent of the interferometer scale factor, and microkelvin temperature LPAI gravimeters with 10~km scale wavefront curvature have biases at the 10~nm/s$^2$ level.

Quantifying the systematic effects of wavefront imperfections on LPAI measurements is a difficult task~\cite{LouchetChauvet2011,Schkolnik2015,zhouObservingEffectWavefront2016,Karcher2018,pagot_influence_2025, luo_evaluating_2025} and often relies on estimations of surface flatness of the optical elements and {\it ex situ} characterization of the interferometer beam. Zero-temperature extrapolation of the wavefront bias has shown that low-order Zernike wavefront analysis is insufficient for bias correction and that the bias can vary non-monotonically with gas temperature~\cite{Karcher2018}. Recently, progress has been made on {\it in situ} wavefront characterization in Ramsey-Borde interferometers, with demonstrations of 1D mapping of the relative integrated atomic phase~\cite{Xu2024} and the k-vector magnitude~\cite{gaudout_probing_2025}, both measured by translating the position of the atoms in the optical field. Mapping of the interferometer phase within a single expanding gas is commonly used in point-source atom interferometry~\cite{Dickerson2013} and has been demonstrated in a Bragg interferometer using principal-component analysis and ellipse fitting~\cite{Seckmeyer2025}.

In this work, we demonstrate {\it in situ} 2D mapping of the interferometer phase in a Raman LPAI to measure and correct the wavefront bias. Imaging of the atoms in the plane transverse to the interferometer's k-vector enables direct measurement of the transverse-motion biases including phases from magnetic gradients and wavefront curvature. To demonstrate accurate measurement of the wavefront bias, we introduce controlled wavefront curvature by varying the collimation of the retro-reflected Raman beam. We combine spatially resolved maps of the interferometer phase with k-reversal~\cite{Durfee2006} to isolate the systematic error due to wavefront distortion, and we demonstrate that the measured wavefront curvature bias can be characterized at the mrad level. 

\begin{figure}
\centering
\includegraphics[width=\linewidth]{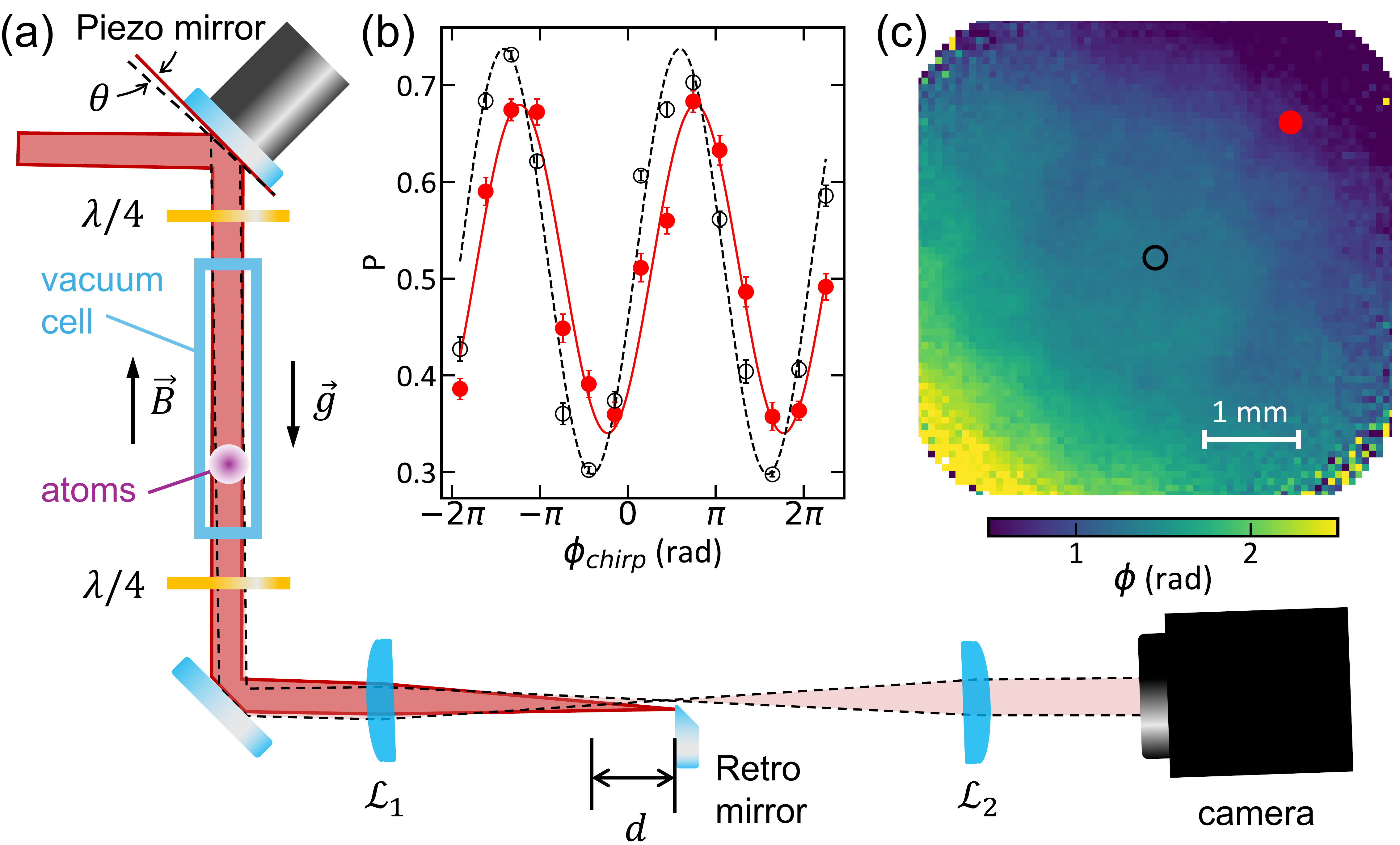}
\caption{{\it In situ} phase mapping. (a) An experimental schematic shows that the Raman interferometer light (red) is retro-reflected using a pick-off mirror placed near the center of a 4-f imaging system. The wavefront curvature of the retro-reflected Raman beam is controlled by translating the retro-reflecting mirror by a distance $d$ away from the focus of lens $\mathcal{L}_1$.  Deflection of this beam path (dashed) allows for imaging in the transverse plane of the interferometer. (b) The local interferometer phase is inferred by spatially measuring the interferometer transfer probabilities as the global interferometer phase is varied. Two characteristic interference fringes are shown (black and red) from the spatial interferometer phase map in (c) for $+k_{\mathrm{eff}}$. Error bars indicate the standard uncertainty of the mean.}
\label{fig:description}
\end{figure}

% \section{Experimental description}
{\it Experimental description}---Spatial wavefront mapping is demonstrated in a Mach-Zehnder ($\pi/2-\pi-\pi/2$) atom interferometer using  laser-cooled $^{87}$Rb atoms in the apparatus described in Ref.~\cite{Chen2019}. Briefly, $\sim 10^7$~atoms are cooled to $\approx$~3~$\mu$K using a 3D~magneto-optical trap and optical molasses cooling. The atoms are allowed to fall under the influence of gravity, and Doppler-sensitive Raman transitions between the $F = 1,2$ ground state levels are used to recoil the atoms by $\hbar k_{\mathrm{eff}}$ as shown in Fig.~\ref{fig:description}(a), where $k_{\mathrm{eff}} =  4\pi/ \lambda$ is the Raman $k$-vector and $\lambda \approx$ 780~nm is the wavelength of the Raman light. 

The interferometer operates with a time $T = 16$~ms between each pulse, and the first Raman pulse is delayed 8~ms after release to resolve the desired Doppler-sensitive transitions. The resonant Raman light fields are detuned by $\approx~210$~MHz below the $F'=3$ excited state and have the same circular polarization~\cite{sm}.  The Raman $\pi$~transition time is $\sim~5\ \mu$s, and the atoms expand for  $T_{\rm ex} \approx 45 \ {\rm ms}$. The experimental drop time, Raman beam size ($\approx 4.5 $ mm $1/e^2$ radius), and $T$ are limited by the cm-scale of the vacuum vessel used in this work. The total experimental cycle is $\sim 1$~s, dominated by the time used to capture atoms in the MOT. 
%Uncompensated vibration noise limits the short term acceleration stability to $\sim 2\times 10^{-5}$ m/s$^2/\sqrt{\tau}$,  where $\tau$ is the measurement averaging time. 

During the interferometer sequence, the two-photon Raman detuning is linearly swept to follow the Doppler shifted resonance as the atoms are accelerated by gravity. The phase of the interferometer is $\phi = (\mathbf{k_{\mathrm{eff}}} \cdot \mathbf{g} - \alpha)T^2+ \phi_{\rm sys}$, where $\mathbf{g}$ is the gravitational acceleration vector, $\alpha \sim \pm \ 2\pi ~25.1 \ {\rm  MHz/s}$ is the Raman chirp rate, and $\phi_{\rm sys}$ are the phases generated from other sources including magnetic fields, light shifts, and Coriolis phases. By changing the sign $\alpha$, we select which pair of counter-propagating Raman fields is resonant and change the sign of the momentum kick imparted to the atoms~\cite{Durfee2006}. The atoms are detected using absorption imaging along the Raman k-vector using 4-f imaging optics $\mathcal{L}_{1,2}$ with focal lengths $f = 150 \ {\rm mm}$. 

The Raman light is retro-reflected in a ``cat-eye" geometry using a pickoff mirror located near the center of the 4-f imaging system. This geometry maintains a constant intensity ratio of the $F = 1,2$ frequency components of the Raman light as there are no polarization sensitive elements. The intensity ratio is chosen to minimize differential light shifts~\cite{petersHighprecisionGravityMeasurements2001a}, and residual light shifts are largely common mode between the $\pm k$ transitions. Imaging is achieved by deflecting this beam path using a piezo-actuated mirror so that the imaging light passes the pickoff mirror and is recorded on a camera. The beam is deflected by $\approx 2.5$~mrad, and this process introduces a 6.8~ms delay between the last Raman pulse and detection.

% \section{Phase mapping}
{\it Phase mapping}---The spatially-resolved interferometer phase is measured using pixel-by-pixel fitting of the interferometer signal~\cite{Hoth2016}. To enable local determination of the interferometer phase $\phi_i$ and contrast $c_i$, the global phase of the interferometer is scanned over a range of $\approx 4\pi$ by varying $\alpha$; the corresponding Raman frequency chirp phase is $\phi_{\textrm{chirp}} = (\alpha - \alpha_0)T^2$, where $\alpha_0$ is the chirp rate which nulls the gravitational phase shift. The spatially-resolved, fractional interferometer transfer probability $P_i$ is fit to a sinusoidal function $P_i= [1 - c_i\cos(\phi_\textrm{chirp}+\phi_i)]/2$ to infer $\phi_{i}$ at each pixel $i$ as shown in Fig.~\ref{fig:description}(b). 

An example of the resulting phase maps is shown in Fig.~\ref{fig:description}(c). The atomic absorption signal is recorded at 9~$\mu$m spatial resolution, and the data is processed after $8\times8$ binning of the data such that each pixel is 72~$\mu$m by 72~$\mu$m. In the center of the gas, each pixel detects $\sim 6000$ atoms and the interferometer contrast is $\approx$~40~\%. For the recorded data, the single-shot phase noise from uncompensated vibrations is $\approx 70$~mrad and dominates the projection and measurement noise except near the edges of the gas.
% The contrast drops to $\approx$~25~\% at the edge of the field of view due to reduced Raman Rabi rates in the wings of the Raman beam.

\begin{figure}
	\centering
	\includegraphics[width=\linewidth]{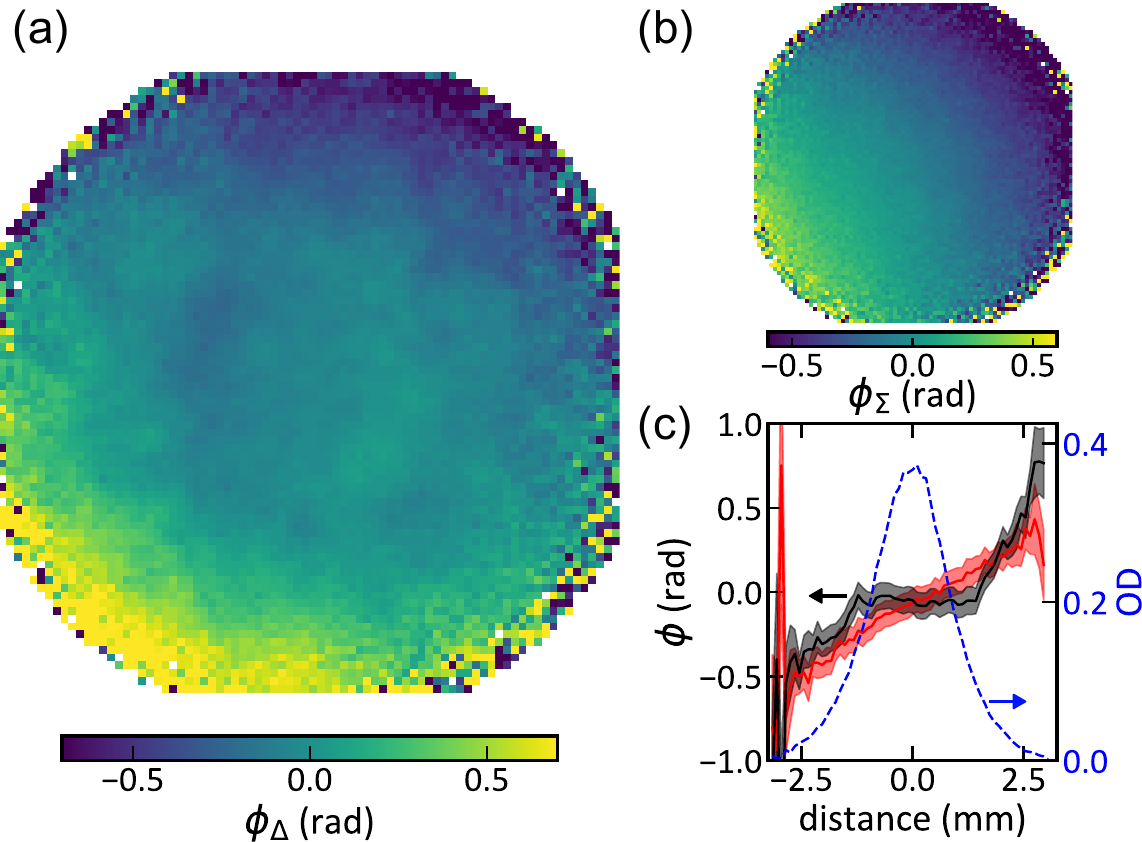}
	\caption{Phase mapping with k-reversal. Maps of $\phi_{\Delta}$ (a) and $\phi_{\Sigma}$ (b) show phases which are differential and common mode under k-reversal, respectively. (c) Traces of $\phi_{\Delta}$ (black), $\phi_{\Sigma}$ (red), and the optical depth (OD) of the gas (blue) are shown along a diagonal of the maps. The field of view in (a,b) is 5.8~mm x 5.8~mm with near zero applied wavefront curvature.}
	\label{fig:fig2}
\end{figure}

Phase variation across the atomic sample is observed at the radian level due to a combination of inertial and non-inertial phases. To aid in isolating the wavefront phases, we perform k-reversal to reject differential light shifts and magnetic gradients that are common-mode to both k-vectors. The $\pm k$ phase maps $\phi_{+k}$ and $\phi_{-k}$ are combined to separate these signals, forming the differential and common-mode signals $\phi_{\Delta} = (\phi_{+k} - \phi_{-k})/2$ and $\phi_{\Sigma} = (\phi_{+k} + \phi_{-k})/2$, respectively, as shown in Fig.~\ref{fig:fig2}.

The differential signal $\phi_{\Delta}$ includes the desired inertial phases, wavefront biases, and other asymmetries which exist in the atom interferometer; a representative image is shown in Fig.~\ref{fig:fig2}(a). The inertial phases include the linear acceleration phase as well as the Coriolis phase gradient $\mathbf{k_\mathbf{\Omega}} = 2(\mathbf{k}_\mathrm{eff}\times\mathbf{\Omega_{\oplus}})T^2/T_\text{ex} \sim 9.7$~rad/m, where $\Omega_{\oplus}$ is the Earth rotation rate~\cite{hoganLightpulseAtomInterferometry2009}. The Coriolis phase has an odd symmetry relative to the transverse velocity, and they do not contribute to the integrated acceleration bias in most LPAIs, except through asymmetrical sampling. Other common asymmetries could bias $\phi_{\Delta}$ including two-photon light shifts~\cite{Gauguet2008}, residual Doppler shifts~\cite{Gillot2016}, and other effects which limit the symmetry of the k-reversal measurement. The inertial phase gradients is small in this measurement, and the spatial variation of $\phi_{\Delta}$ arises primarily from wavefront aberrations. 

\begin{figure}
	\centering
	\includegraphics[width=\linewidth]{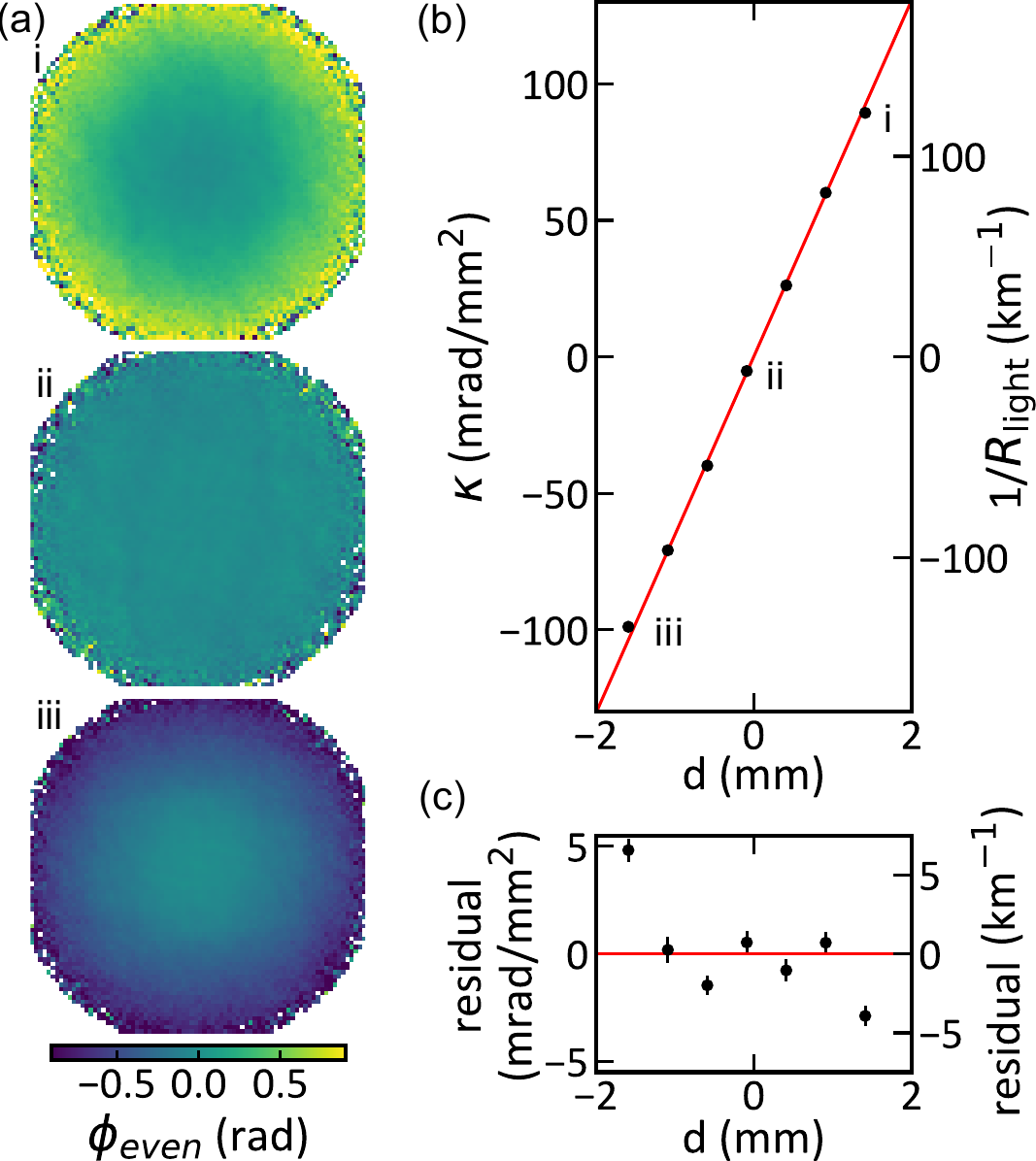}
	\caption{Wavefront curvature measurement. (a) Maps of $\phi_{\rm even}$ are recorded at varying Raman wavefront curvature set by the retro-reflecting mirror position $d$. (b) The curvature of $\phi_{\mathrm{even}}$ is extracted from these maps (black points) and compared to the predicted curvature from Eq.~(\ref{equ:radius correction}) (red line). Labeled points correspond to the maps in (a) and residuals are shown in (c). The right axes in (b,c) are inferred using Eq.~\ref{equ:radius correction}; some error bars are smaller than the markers.}
	\label{fig:fig3}
\end{figure}

The measured $\phi_{\Delta}$ have a strong anti-symmetry around the center of the transverse velocity distribution, as visible in Fig.~\ref{fig:fig2}(a), arising from the $-1$ magnification ratio of the ``cat-eye" retro-reflector. In this geometry, any wavefront structure in the incident Raman laser beam is spatially inverted upon retro-reflection and contributes to the odd symmetry of the $\phi_{\Delta}$ maps. The odd components of  $\phi_{\Delta}$ do not bias the integrated phase measurement used in gravimetry but would bias measurements made in point-source atom gyroscopes~\cite{Dickerson2013,Hoth2016}. We isolate the even, phase-biasing signal by symmetrizing the images according to $\phi_{\rm even}(\mathbf{r}) = \left(\phi_{\Delta}(\mathbf{r}) + \phi_{\Delta}(-\mathbf{r})\right)/2$, where $\mathbf{r}$ is the position vector relative to the center of the final gas; example of these maps are shown in Fig.~\ref{fig:fig3}. 
% The even components of $\phi_{\Delta}$ do generate bias for acceleration measurements.

The common mode signal $\phi_{\Sigma}$ shown in Fig.~\ref{fig:fig2}(b) includes effects which do not depend on the sign of the k-vector. Here, $\phi_{\Sigma}$ is dominated by a magnetic gradient estimated at $\approx 0.6 ~\mu{\rm T/mm}$. This gradient generates an interferometer phase gradient $\phi^{\prime} \approx -2\gamma B |B|^{\prime} T^2/ T_{ex}$ where  $\gamma$ is the 2nd-order Zeeman shift coefficient, $B \approx 70 \mu{\rm T}$ is the field strength, and $|B|^{\prime}$ is the field's spatial gradient in the transverse plane. Magnetic gradients along the Raman k-vector are not resolved in this signal and could contribute to $\phi_{\Delta}$ since the atoms travel on slightly different trajectories for $\pm k$. 

% \section{Wavefront characterization}

{\it Wavefront characterization}---To demonstrate wavefront characterization, we vary the Raman wavefront curvature and measure the impact on $\phi_{\mathrm{even}}$ as shown in Fig.~\ref{fig:fig3}(a). The curvature is introduced by controllably varying the collimation of the retro-reflected Raman light by translating the pickoff mirror within the ``cat-eye" retro-reflector as indicated in Fig.~\ref{fig:description}(a). This generates a Raman wavefront difference with a radius-of-curvature $R_{\mathrm{light}}=f^2/2d$, where $d$ is the displacement of the mirror from the focal plane of lens $\mathcal{L}_1$.  

The applied wavefront curvature modifies the interferometer phase as atoms sample different points in the Raman beam at each of the LPAI light pulses. A useful approximation for the wavefront phase in an LPAI is $\phi_{\mathrm{atom}} = \phi_{\mathrm{light}}(\mathbf{r}_1) - 2\phi_{\mathrm{light}}(\mathbf{r}_2) + \phi_{\mathrm{light}}(\mathbf{r}_3)$, where $\phi_{\rm light}$ is the phase of Raman light and $\mathbf{r}_i$ is the location of the atom at the $i$-th light pulse~\cite{hoganLightpulseAtomInterferometry2009}.  For the case of a parabolic wavefront, such as the kind generated by varying the collimation of the Raman light, $\phi_{\mathrm{light}} =  k_{\mathrm{eff}}\mathbf{r}^2/ 4R_{\mathrm{light}}$ and  $\phi_{\mathrm{atom}}(\mathbf{v}) =  k_{\mathrm{eff}} \mathbf{v}^2T^2 / 2 R_{\mathrm{light}}$~\cite{weissPrecisionMeasurementMCs1994}, where $\mathbf{v}$ is an atom's transverse velocity vector. 
% This yields an expected parabolic phase of the form $\phi_{\mathrm{even}} = \kappa_R \mathbf{r}^2$. 

Practical LPAIs have a spread of velocities at each $\mathbf{r}$ due to the finite size of the initial, laser-cooled gas, and the observed wavefront curvature is reduced by the imperfect position-velocity correlation observed at the time of imaging. For a thermal gas with a Gaussian initial size $\sigma_0$ and final size $\sigma_f$, the expected wavefront is
\begin{equation}\label{equ:radius correction}
\phi_{\rm even}(\mathbf{r})=\frac{k_\text{eff}T^2}{2R_{\mathrm{light}}T_\text{ex}^2} \left(\beta^2 \mathbf{r}^2 + 2\beta\sigma_0^2\right),
\end{equation}
where $\beta = 1-\sigma_0^2/\sigma_f^2$~\cite{sm}. Equation (\ref{equ:radius correction}) is calculated by integrating the wavefront phase shifts over the atomic velocity and initial position distributions, similar to the calculation of finite-size corrections in point-source atom interferometry~\cite{hothDevelopmentCharacterizationInterferometer2015}. Equation (\ref{equ:radius correction}) includes a phase offset $\phi^*$ at $\mathbf{r} = 0$; this offset is maximally $\phi^* = k_{\rm eff}\sigma_0^2/8R_{\rm light}$ in the limit of large expansion where $\beta = 1$ and $T_{\rm ex } = 2T$.

In our LPAI, the spatial phase mapping procedure with k-reversal was repeated as $R_{\mathrm{light}}$ was varied over a range of $\approx -8$~m to $+8$~m as shown in Fig.~\ref{fig:fig3}. For these measurements, 480 runs of the interferometer sequence were used at each mirror position, and the maximum variation of the retro-reflected light intensity from the applied curvature is $< 10 \ \%$ at the atoms. The resulting maps of $\phi_{\mathrm{even}}$ were fit to parabolic surfaces~\cite{sm} as 
\begin{equation}\label{equ:phioffset}
\phi_{\rm even} = \kappa \mathbf{r}^2 + \phi_{\mathrm{offset}}.
\end{equation}
The extracted values for $\kappa$ are shown in Fig.~\ref{fig:fig3}(b), and the data agree with Eq.~(\ref{equ:radius correction}) with no free parameters as shown in the residuals plotted in Fig.~\ref{fig:fig3}(c). The ratio $\sigma_0/\sigma_f \approx 0.33$ for this data. 

\begin{figure}
\centering
\includegraphics[width=0.9\linewidth]{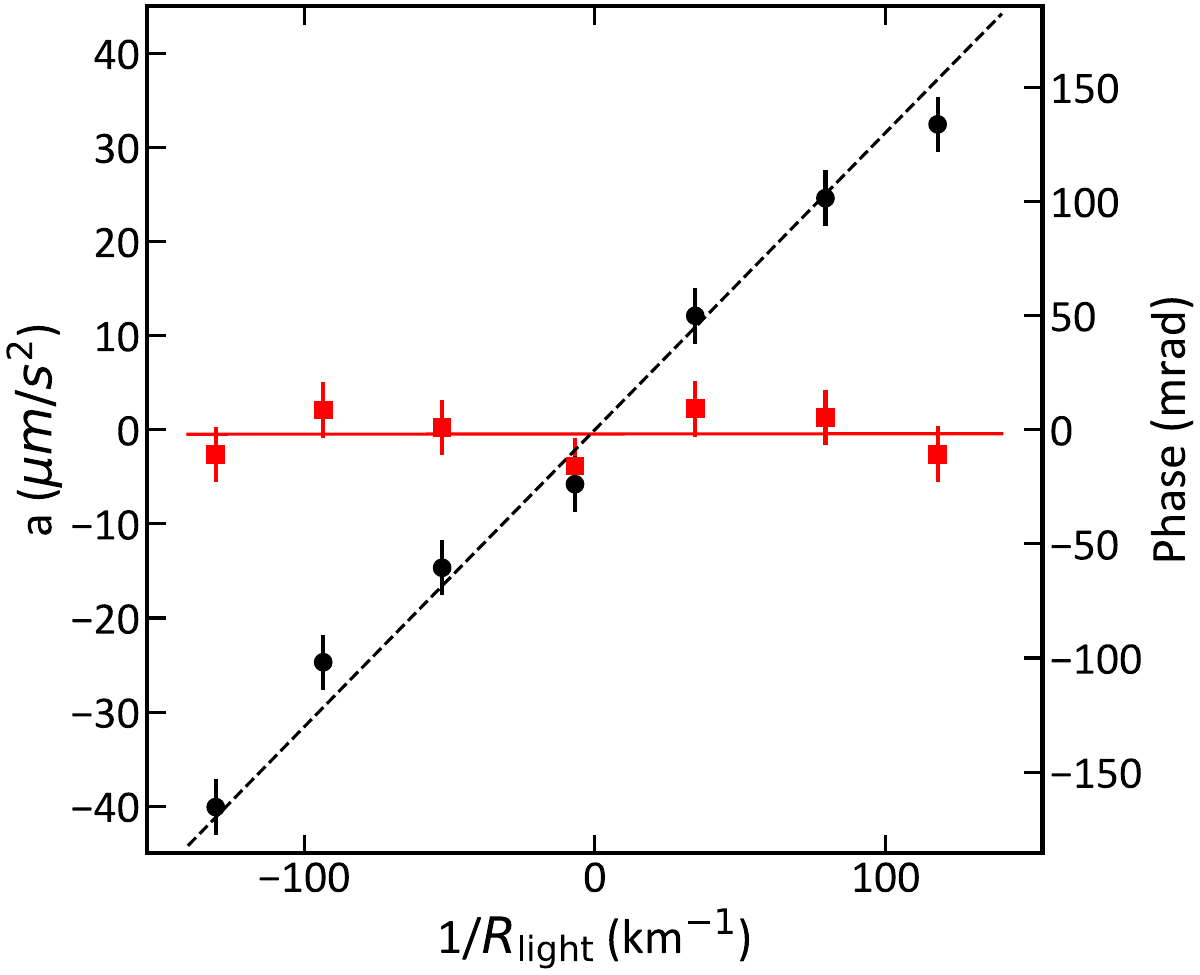}
\caption{Wavefront bias and correction. The integrated acceleration signal (black circles) is shown as the wavefront curvature is varied (same dataset as Fig.~\ref{fig:fig3}) and compared to the expected bias $\left<a_{\mathrm{bias}}\right> = \sigma_v ^2/R_{\mathrm{light}}$ (dashed black line). A bias-free signal (red squares) is extracted from the central ($|\mathbf{r}| < \sigma_0$) interferometer phase accounting for the finite size bias $\phi^*$. A linear fit to the bias-corrected data (solid red line) is consistent with zero slope, implying strong suppression of wavefront curvature biases. The datasets are offset so that the $a$ is approximately zero at $R_{\rm light}^{-1} = 0$.}
\label{fig:fig4}
\end{figure}

% \section{Bias correction}
{\it Bias correction}---The effect of wavefront curvature on the measured acceleration signal is quantified using the data in Fig.~\ref{fig:fig3}. The average acceleration across the measured $\phi_{\Delta}$ maps is calculated as 
\begin{equation}\label{equ:numbias}
\left<a\right>  = k_{\rm eff}T^2 < \phi_{\Delta,i} > , 
\end{equation}
where the brackets indicate a weighted averaging over the phases measured at each pixel $i$. The weights are set by the occupation of each pixel $N_i$, analogous to the averaging that occurs in integrated fluorescence detection in conventional LPAIs. The results are shown in Fig.~\ref{fig:fig4}, where the data are plotted against $R^{-1}_{\mathrm{light}}$ from Fig.~\ref{fig:fig3}. 

The measured acceleration depends linearly on the inverse of the wavefront curvature~\cite{petersHighprecisionGravityMeasurements2001a}, and the values are consistent with the expected bias for a thermal gas as
\begin{equation}\label{equ:avgbias}
\left<a_{\mathrm{bias}}\right>  = \sigma_v^2/R_{\mathrm{light}}, 
\end{equation}
where $\sigma_v = \sqrt{k_B \tau/m}$, $\tau = (3.3 \pm 0.2)\  \mu$K is the gas temperature, and $m$ is the atomic mass. The acceleration bias is independent of the interferometer time $T$ as the scale factor and phase bias both scale with $T^2$.  The bias is independent of finite size effects, but does depend on the distribution of atomic velocities. The uncertainty of the bias characterization is set by the measurement uncertainty of $R_{\rm light}$ and of the gas temperature. For the measurement in Fig.~\ref{fig:fig3}(ii), $R^{-1}_{\rm light} = (-6.83 \pm 0.68) \ {\rm km}^{-1}$ and the acceleration bias is $(-21.5 \pm 2.5)\times 10^{-7} \ {\rm m/s^2}$, equivalent to $(-8.9 \pm 1.0)$~mrad in phase bias. 

A ``bias-free" inertial signal can also be extracted from $\phi_{\rm even}$ by evaluating the central  phase $\phi_{\rm even}(\mathbf{r}\!=\!0)$ and subtracting the finite-size phase offset $\phi^*$ as in Eq.~(\ref{equ:radius correction}). The central phase can be extracted using surface fitting of the phase maps, but we evaluated this phase numerically using the central 10~\% of the atoms and the weighting used in Eq.~(\ref{equ:numbias}) to avoid potential artifacts from higher-order wavefront curvature. Here, the phase offset $\phi^* = \beta \sigma_0^2 k_{\rm eff} (T/T_{\rm ex})^2/R_{\rm light} \approx 150 \ {\rm \mu rad \ km}$, equivalent to $\approx 36 \ {\rm nm/s^2 \ km }$ for T = 16 ms. The resulting inertial signal $a = k_{\rm eff} T^2 (\phi_{\rm even}(\mathbf{r}\!=\!0) - \phi^*)$ rejects the wavefront curvature bias and is shown as the red squares in Fig.~\ref{fig:fig4}. A line fit to this data of form $a = \alpha  R^{-1}_{\textrm{light}}$ produces a slope of $\alpha = (0.2 \pm 13)$~nm/s$^2$~km, consistent with zero. Whereas, the sensitivity of the uncorrected acceleration signal to wavefront curvature from Eq.~(\ref{equ:avgbias}) is $\alpha = \sigma_v^2\approx 290 \ {\rm nm/s^2 \ km}$.

The uncertainty of the corrected phases in Fig.~\ref{fig:fig4} are dominated by laboratory vibration noise to $\approx 10 \ {\rm mrad}$ at 8 minutes of integration time, and the mrad-level wavefront bias uncertainty exceeds the stability of the interferometer signal. This level of phase bias uncertainty would allow for $\approx 5 \ {\rm nm/s^2}$ wavefront bias uncertainty at $T = 100$~ms, which could be reduced below $1 \ {\rm nm/s^2}$ using longer averaging times and vibration rejection. The quantum projection noise (QPN) limit \cite{itanoQuantumProjectionNoise1993} to wavefront curvature measurement is calculated using a least-squares approach~\cite{liSensitivityPointSourceInterferometryBasedInertial2024} as $\sigma_{R_{\rm light}^{-1}}  = 1/k_{\rm eff}T^2\sqrt{2N}\sigma_v^2$, where $N$ is the total number of atom in the interferometer. The QPN limit on the associated wavefront bias uncertainty is 
\begin{equation}\label{equ:final}
\sigma_{<a_{\rm bias}>} = \frac{d<a_{\rm bias}>}{ d R_{\rm light}^{-1}} \sigma_{R_{\rm light}^{-1}} = \frac{1}{\sqrt{2N}k_{\rm eff}T^2},
\end{equation}
which is independent of $R_{\rm light}$ and similar to the QPN-limited acceleration uncertainty $\sigma_a = 1/\sqrt{N}k_{\rm eff}T^2$~\cite{sm}. This implies that wavefront systematics could be evaluated at the level needed to reduce absolute acceleration uncertainty in leading atom interferometers. 

% \section{Conclusion}
{\it Conclusion}---Transverse spatial phase mapping in LPAI, as demonstrated in this work, enables {\it in situ} evaluation of wavefront biases  and reduction of the absolute uncertainty of the inertial atom interferometer phase. This approach is complimentary to methods that strongly suppress the gas temperature and associated transverse motion biases using evaporative cooling or $\delta$-kick cooling, which may not be feasible in compact or transportable atom interferometers. Here, we show that evaluation of the parabolic wavefront phase at the $\approx$ 1.0~mrad level, which would enable optimized atom interferometers to reduce the wavefront bias from the 30 nm/s$^2$ level to below 1~nm/s$^2$.  

This study focused on the parabolic wavefront bias, but higher-order wavefront analysis is likely necessary to capture the full absolute bias on inertial phases for more complex wavefront profiles ~\cite{Gillot2016}. This analysis would involve fitting of $\phi_{\Delta}$ to Zernike polynomials and involve corrections to the gravimetric phase for each of the even-order terms. The higher order terms do not have constant curvature, and measuring the final position of the atoms in the gas with respect to the wavefront would be necessary to correct these biases. We have performed this Zernike analysis up to 8th-order for our measurements of $\phi_{\Delta}$, and the resulting curvature values are consistent with the $\phi_{\mathrm{even}}$ analysis presented in Fig.~\ref{fig:fig3}. 

Transverse phase mapping is a direct method for measuring the wavefront biases and could be incorporated into many LPAI systems as a calibration step or as the primary phase measurement. The ``cat-eye" retro-reflection approach used here enabled imaging in the transverse plane and provided a convenient method to control the curvature of the wavefront, but other methods for retro-reflection may be more practical in fieldable gravimeters and large-scale interferometers, including use of dichroic optics or partially-reflective mirrors, without the need for a rapid mirror motion. Spatial wavefront mapping may prove useful in a broad range of LPAI systems including gyroscopes, long-baseline interferometers~\cite{abeMatterwaveAtomicGradiometer2021}, and other measurements where mrad-level phase accuracy is needed~\cite{hoganAtomicGravitationalWave2011b}.  

% \section{Acknowledgment}
{\it Acknowledgment}---This work was supported by the National Institute of Standards and Technology and the ``NIST on a Chip" program. JJ acknowledges the financial assistance award 70NANB23H027 from U.S. Department of Commerce, NIST. We thank Greg Hoth and Stephen Eckel for helpful suggestions on this manuscript, and we thank Elizabeth Donley for her years of support as chief of the Time and Frequency Division at NIST. 

{\it Data Availability}---The data that support the findings of this article are not publicly available upon publication because it is not technically feasible and/or the cost of preparing, depositing, and hosting the data would be prohibitive within the terms of this research project. The data are available from the authors upon reasonable request.

\bibliography{Biblio}

\end{document}